          [2001/04/25 1.1 (PWD)]
\documentclass[a4paper,twocolumn]{esapub} 
\usepackage{natbib}
\usepackage{epsf,psfig,epsfig}

\title{Strong QPOs and High Energy Tail in Simultaneous RXTE/INTEGRAL Observations of GRS~1915+105}
\author[1,2]{J\'er\^ome Rodriguez}
\affil[1]{CEA Saclay, Service D'Astrophysique, France}
\affil[2]{ISDC Versoix, Switzerland}
\author[1]{Ya$\ddot{e}$l Fuchs}
\author[3]{Diana Hannikainen}
\author[3,2]{Osmi Vilhu}
\affil[3]{Observatory of Helsinki, Finland}
\author[4,2]{Simon Shaw}
\affil[4]{University of Southampton, the UK}
\author[5]{Tomaso Belloni}
\affil[5]{Osservatorio Astronomico di Brera Itlay}
\author[6,1]{St\'ephane Corbel}
\affil[6]{Universit\'e Paris 7, France}

\begin{document}


\maketitle

\begin{abstract}
We present the first results of the timing analysis of our {\emph{RXTE/INTEGRAL}} monitoring 
campaign on GRS~1915+105. Over the 6 already performed 
{\emph{RXTE}} observations, we study the presence of Low Frequency QPO (LFQPO),  and 
their energetic dependences. In a view to understand the QPO phenomenon, 
we compare the QPO properties to the spectral behaviour of the source. We propose 
that part of the compact jet detected during multi-wavelength observations, 
could produce a significant amount of hard X-rays, and hence explain the 
energy dependence of the amplitude of the QPOs.
\end{abstract}

\section{Monitoring campaign on GRS~1915+105}
\begin{table*}
\begin{tabular}{|ccccc|}
\hline
Obs. sequence \# & Date (MJD)&  Good times &  Revolution  \# & {\emph{INTEGRAL}} ref. \\
({\emph{RXTE}}) & & (s) & ({\emph{INTEGRAL}}) & \\
\hline
1 & 6-7 March 2003 (52704-05)&15768 & Rev. 48 & Hannikainen et al. 2003 \\
2 & 2 April (52731)&9300 & Rev. 57 & Fuchs et al. 2003 \\
3 & 9-10 April (52738-39)&25360 & Rev. 59 & Hannikainen et al. 2004 \\
4 & 9 May (52768)& 14000& Rev. 69 & Hannikainen et al. 2004 \\
5 & 2 November (52945) & & Rev. 122 & Solar Flares\\
6 & 22-23 November(52965-66) &36100& Rev 135 & Hannikainen et al. 2004 \\
\hline
\end{tabular}
\caption{log of the {\emph{RXTE}} observation and contemporaneous {\emph{INTEGRAL}} revolutions reported in
this paper}
\label{tab:log}
\end{table*}
The log of the {\emph{RXTE}} observations analysed in this paper is reported in 
Table \ref{tab:log}. All of them were simultaneous with {\emph{INTEGRAL}} and other ground 
based observations. The first observation shows a new class of variability 
(Hannikainen et al. 2003), while during Obs. 2,3,4, the source has a steady flux 
in the X/Gamma rays. It shows a strong QPO, a  powerful compact jet during Obs. 2 
(Fuchs et al. 2003), and a high level of radio emission during the 2 other 
observations (Hannikainen et al. 2004, these proceedings).

\section{Class of variability over the campaign}
The study of the {\emph{RXTE}} colour-colour diagrams allowed us to classify the class
of variability of the source following Belloni et al. (2000) classification:
\begin{itemize}
\item Observation \#1: as shown in Hannikainen et al. (2003) and discussed in these 
proceedings, this observation belongs to a new class of variability
\item Observations \#2,3,4: these three observations belong to the steady class $\chi$.
The high level of radio emission allows to further classify them as $\chi1-\chi3$, a.k.a 
radio loud hard state (Muno et al. 2001), or type II state (Trudolyubov 2001). 
See Fuchs et al. (2003) for a presentation of the whole multi-wavelength campaign 
during Obs. 2.
\item Observation \#5 : a high level of solar activity renders the analysis of this 
observation delicate. It will be presented elsewhere 
\item Observations \#6: this last observation of our {\emph{RXTE}} AO-8 campaign belongs to the steady class 
$\phi$. 
\end{itemize}


\section{Class $\chi$ preliminary spectral analysis}
\begin{figure}[htbp]
\epsfig{file=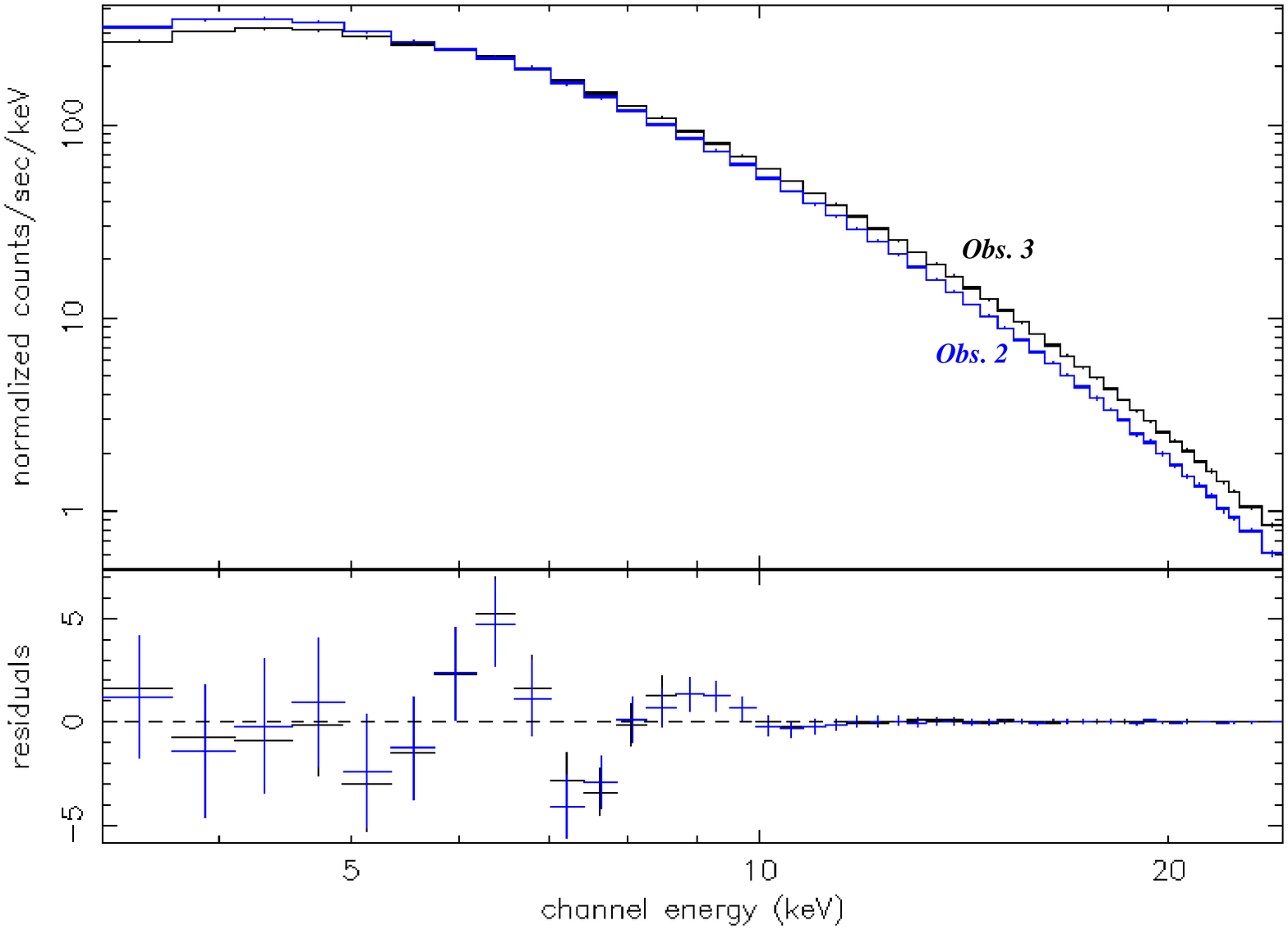,width=\columnwidth}
\caption{PCA 2--25 keV spectra from Obs. 2  and Obs. 3, with the best fit
model superimposed.}
\label{fig:PCA}
\end{figure}

Fits of the PCA 2-25 keV spectra with a standard model of (absorbed)  
multi colour disc black body 
plus power-law leads to unrealistic values of the disk parameters (kT $\sim$4 keV)
as already reported in the literature for such radio loud states (e.g. Muno et al. 2001). 
When restraining to the 3-25 keV (PCA only), a cut-off power-law (with $\Gamma$=1.8) 
fits  the spectra as shown in Fig. \ref{fig:PCA} for both Obs. 2 and Obs. 3.

\begin{figure*}
\begin{tabular}{cc}
\epsfig{file=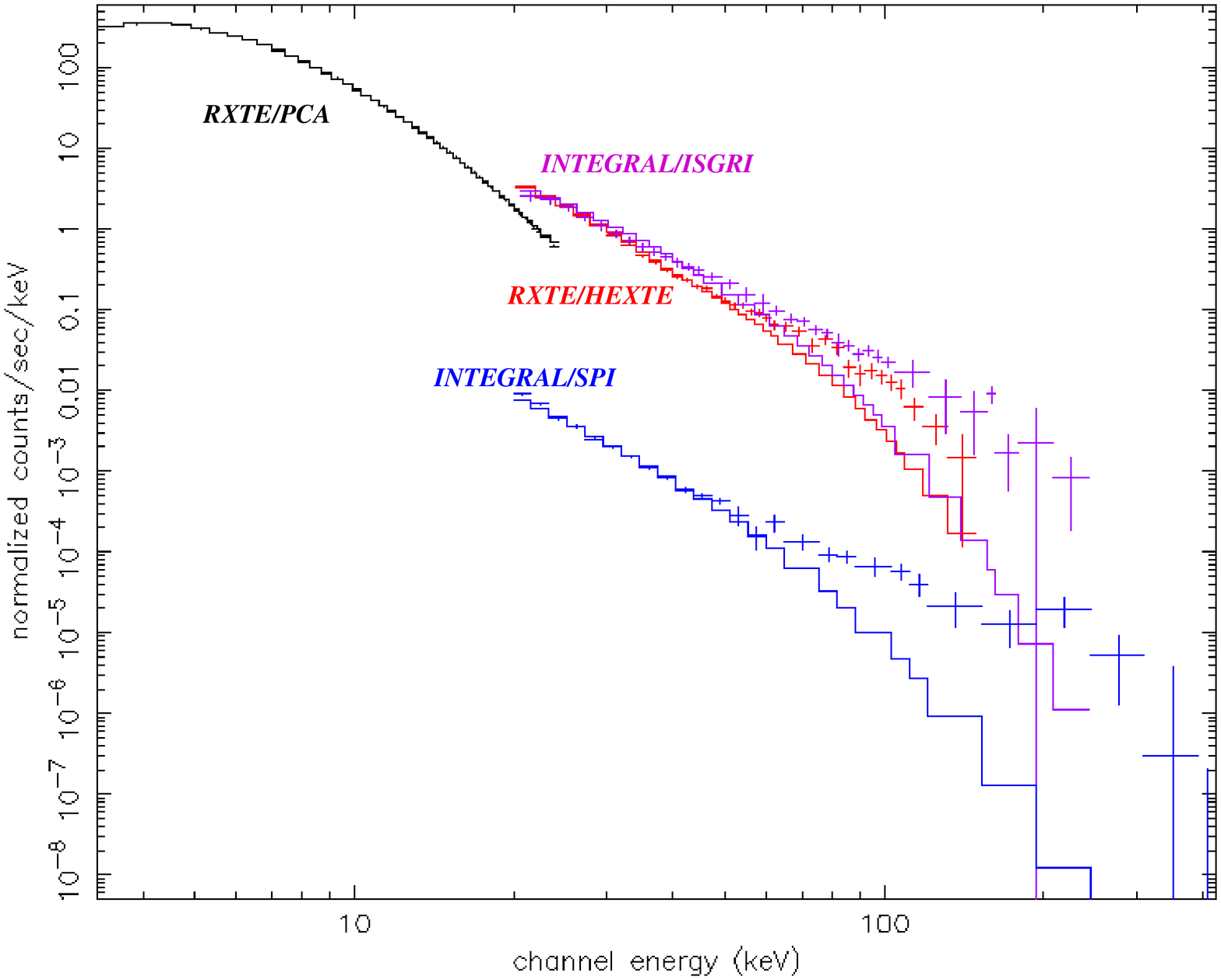,width=8cm}&
\epsfig{file=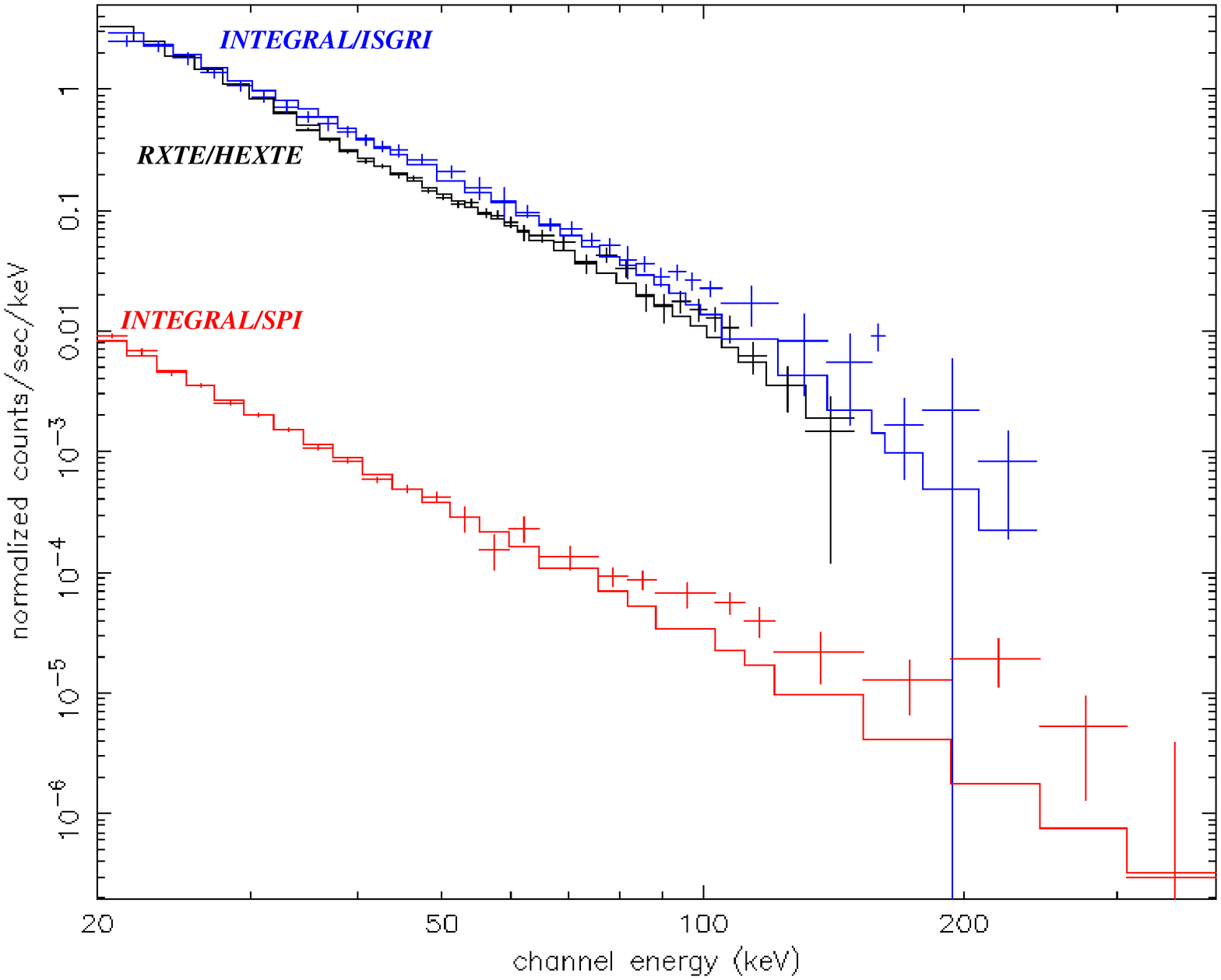,width=8cm}\\
\end{tabular}
\caption{Left: Simultaneous fitting of {\emph{RXTE}} (PCA+HEXTE) and {\emph{INTEGRAL}} (ISGRI+SPI). with a simple 
model consisting of (absorption) multi colour disc black body+cutoff power-law. The deviation at high energy is clear.
Right: High energy (HEXTE+ISGRI+SPI) fit with a simple power-law with $\Gamma\sim3.5$}
\label{fig:High_E}
\end{figure*}

The disc parameters are then closer to what is usually observed in other black hole systems 
(kT~0.5-0.8 keV depending on N$_H$).
This model does not fit  the spectra well when including energies above 25 keV (Fig. \ref{fig:High_E}
 left panel). 
The 20-300 keV spectra are well fitted by a power-law with $\Gamma\sim$3.5 (Fig. \ref{fig:High_E} 
right panel).
A broken power law  with a break energy of $\sim$15 keV ($\Gamma_1$=2.5 $\Gamma_2$=3.5), is the 
best model for the broad band spectrum.  Whether the second power-law is evidence or not 
for a 3rd emitting media during radio loud hard state is thus an open question 
(as already pointed out by Trudolyubov 2001)

\section{Timing Analysis: LFQPOs}
\begin{figure*}
\begin{tabular}{cc}
\epsfig{file=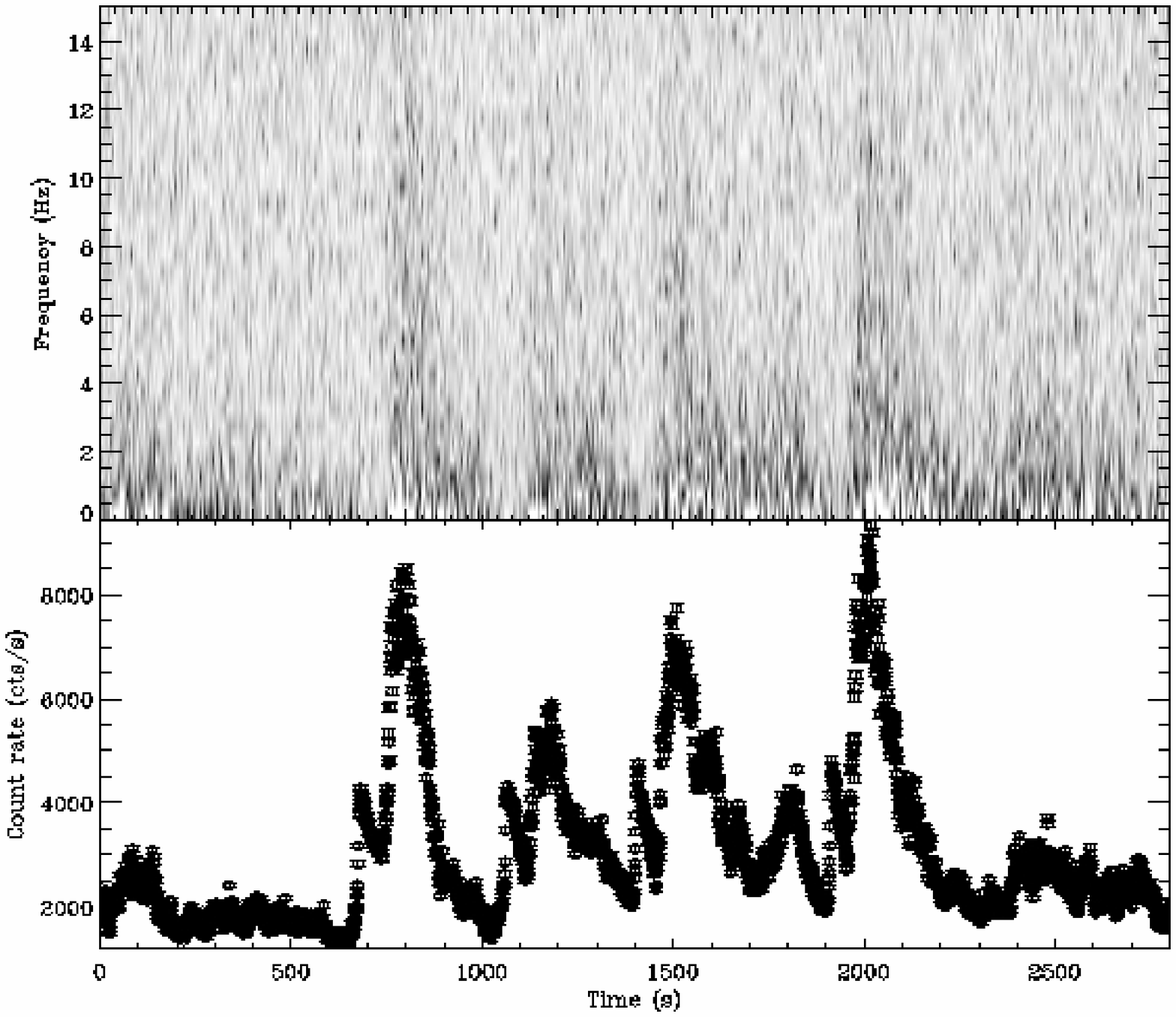,width=8cm}&
\epsfig{file=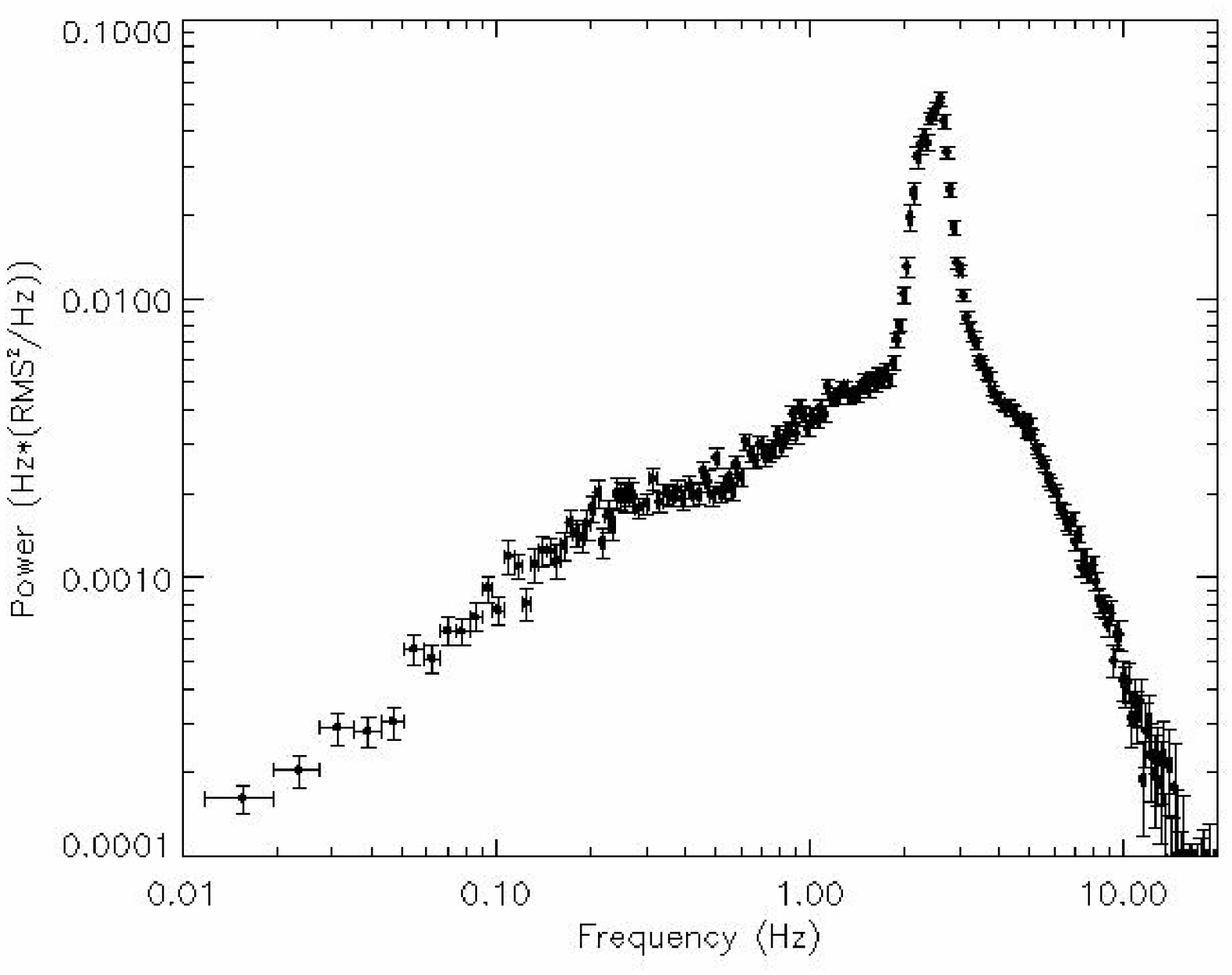,width=8cm}\\
\end{tabular}
\caption{Left: Dynamical power spectrum of a sample of Obs. 1.
Right: power spectrum of Obs. 2 showing the strong QPO. Obs. 3 and Obs. 4 have similar power spectra, although 
in the latter case two different features re found (at different times).}
\label{fig:qpo}
\end{figure*}

\begin{figure}
\epsfig{file=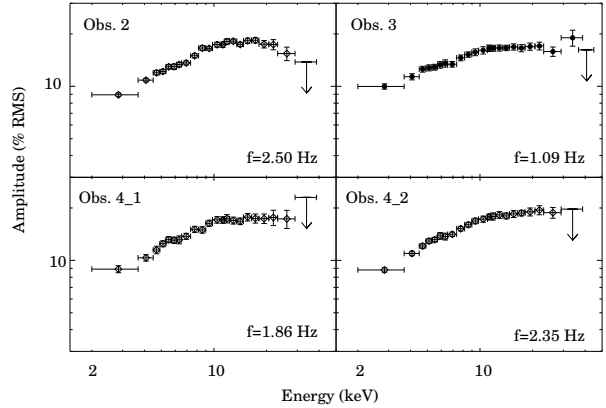,width=\columnwidth}
\caption{Energy dependence of the QPO amplitude for the 4 steady QPOs discussed in the text.}
\label{fig:amplitude}
\end{figure}

In Obs. 1 a  QPO feature is clearly detected in the dynamical power spectra (Fig. \ref{fig:qpo}). 
Its frequency is clearly correlated to the PCA flux as widely reported in the literature 
for  such features (e.g. Markwardt et al. 1999; Rodriguez et al. 2002). 
In Obs. 2,3,4 a steady QPO is clearly detected in the power spectra (e.g. Fig. \ref{fig:qpo} right). 
The feature has a constant frequency over the whole observing time during Obs. 2 and 3. In Obs. 4 
however two distinct frequencies are identified. Due to long exposure we could study the 
energy dependence of the QPO amplitude with the highest  possible accuracy (Fig. \ref{fig:amplitude}).

Although the three observations show that the source is in the same class of variability, the energy 
dependence of the QPO amplitude is quite different from one observation to the other. The main 
difference is the presence of a turn-over in the amplitude vs. energy relation in 
Obs. 2 which is not obvious in the other ones (although a flattening is always detected).
Obviously, this is not related to the frequency of the feature: e.g. the 2 features 
observed in Obs. 4 have a similar energy dependence of their amplitude, 
although they have different frequencies.   

\section{Discussion}
We found that the radio loud hard state (a.k.a. Class $\chi_1$-$\chi_3$) does not lead to 
sensible parameters when fitted with a multi colour disc black body and a power-law (see Muno et al. 2001, 
Trudolyubov 2001) .  A broken power-law  represents
the 3-300 keV spectra well. \\
\indent	During the 3 steady observations high level of radio emission is observed 
(see  Fuchs et al. 2003, Hannikainen et al. these proceedings).  The hard X-rays 
above 20 keV may originate from the jet (as expected see Markoff et al. 2003; Corbel et al. 2003). 
The spectra of GRS 1915+105, would then fit better in the standard picture of micro-quasars states.\\
\indent	Our analysis of the spectra of LFQPOs confirms the presence of a cut-off in their 
energy dependence (Tomsick \& Kaaret 2001; Rodriguez et al. 2002), with an evolving 
energy from $\sim$15-20 keV in Obs. 2 to a value $ >$25 keV in the remaining Obs. 
(Fig.\ref{fig:amplitude}). 
Although  the QPO cut-off is needed at some point (otherwise its amplitude would grow indefinitely),
 its evolution is unclear. It can be easily understood, however,
  if the jet contributes significantly to the hard X-ray, and its flux is not modulated on such 
a short time scale.  \\
\indent	The fact  that a high level of radio emission (at 15 GHz) is found during obs.2, 
 when the cut-off is clearly detected in the QPO spectrum, is compatible with this interpretation. 
We thus propose that a part of the hard X-ray is emitted by the compact jet, explaining 
both the spectral behaviour of the source and the QPO ``spectra''.
  
\section*{Acknowledgements}
JR and YF acknowledge financial support from the French Space Agency (CNES). DH is a Fellow of 
the Finnish Academy. 
JR would like to thank V. Beckmann, C. Cabanac, S. Chaty, J. Chenevez, L. Foschini, A. Gros, 
A. Goldwurm, E. Kuulkers, N. Lund, J. Malzac, A. Paizis, P.O. Petrucci, J.A. Tomsick for very
useful and fruitful discussions all along the duration of the workshop.

\end{document}